`Computer Science`

# Research Report

SAFE DEALS BETWEEN STRANGERS

Henry M. Gladney

IBM Almaden Research Center
San Jose, California 95120-6099
Internet: gladney@almaden.ibm.com

```
Computer Science
```

# Safe Deals between Strangers

Henry M. Gladney

IBM Almaden Research Center
San Jose, California 95120-6099
Internet: gladney@almaden.ibm.com

**Abstract:**
E-business, information serving, and ubiquitous computing will create heavy request traffic from strangers or even incognitos.  Such requests must be managed automatically.  Two ways of doing this are well known:  giving every incognito consumer the same treatment, and rendering service in return for money.  However, different behavior will be often wanted, e.g., for a university library with different access policies for undergraduates, graduate students, faculty, alumni, citizens of the same state, and everyone else.

For a data or process server contacted by client machines on behalf of users not previously known, we show how to provide reliable automatic access administration conforming to service agreements.  Implementations scale well from very small collections of consumers and producers to immense client/server networks.  Servers can deliver information, effect state changes, and control external equipment.

Consumer privacy is easily addressed by the same protocol.  We support consumer privacy, but allow servers to deny their resources to incognitos.  A protocol variant even protects against statistical attacks by consortia of service organizations.

One e-commerce application would put the consumer's tokens on a smart card whose readers are in vending kiosks.  In e-business we can simplify supply chain administration.  Our method can also be used in sensitive networks without introducing new security loopholes.

# CONTENTS

**Introduction**

The public has in recent years begun to accept the promise of an information infrastructure of vast dimensions. Access to network resources will become almost as easy as to local digital resources. Request traffic is likely to grow rapidly from today's already impressive levels. More access to more resources suggests the need for more care than ever before to ensure that valuables are used only as proper authorities wish. The digital security community is responding vigorously, but as yet has not invented methods that work well across the boundaries that separate one enterprise from another. We address two parts of this: authorizing outside users to valuable resources and protecting users' privacy.

For ubiquitous computing (computers embedded in appliances and other machinery) and e-business applications, unregistered users will often request privileged services. Their requests must be quickly accepted or rejected without this action requiring human help. Two effective policies are well known: giving every consumer the same treatment, and giving the service in return for money. However, many services and many consumers want different behavior. This has been identified as a difficult and urgent problem [Wiseman 1998]; also

> "The basic cross-organizational access management problem is exemplified by most licensing agreements for networked information resources today; it also arises in situations where institutions agree to share limited-access resources with institutions as part of consortia or other resource sharing collaborations. In such an agreement, an institution—a university, a school, a public library, a Corporation—defines a user community which has access to some network resource. This community is typically large, numbering perhaps in the tens of thousands of individuals, and membership may be volatile over time, reflecting for example the characteristics of a student body. The operator of the network resource, which may a web site, or a resource reached by other protocols such as Telnet terminal emulation or the Z39.50 information retrieval protocol, needs to decide whether users seeking access to the resource are actually members of the user community that the licensee institution defined as part of the license agreement."
>
> [Lynch 1998]

We now communicate an elegant solution. For a data or process server contacted by client machines on behalf of users not previously known, we show how to enforce service agreements automatically. Implementations will scale well from very small collections of consumers and producers (servers) to immense client/server networks. Servers are not limited to delivering information, but may effect state changes (database changes or controlling machinery).

We exploit the facts that the participants in many transactions need only limited mutual trust, that consumers and services often have organizational affiliations, and that consumer and producer organizations often enter into service agreements that can be represented by digital tokens. We describe protocols for managing, communicating, and using such tokens. End users can choose to obscure their identities, revealing only as much about themselves as is needed to obtain the service wanted at the moment. Well-known digital signature techniques prevent unauthorized sharing of authorization tokens.

Recall that in common business situations certain pairs of people are mutually trusting and trustworthy. Consider emergency road service negotiated by automobile associations such as the AAA with their members and with road service companies respectively. The pairs who trust each other are:

- Consumers and (administrators of) consumer organizations (e.g., automobile club members and the AAA); consumers *enroll* in consumer organizations, and typically hold *enrollment* certificates, which are often differentiated even when issued by a single organization (e.g., "platinum" credit cards);

- Producers and producer organizations (e.g., emergency vehicle drivers and their dispatching agent, ACME Road Services); each kind of service is typically associated with a service *ticket*, which might be used by a producer to claim compensation;
- (Administrators of) consumer organizations and (administrators of) producer organizations (e.g., AAA and ACME respectively), that trust each other just enough to negotiate service agreements; a service agreement can be specified as a map from enrollments to tickets.

These three trust relationships are sufficient for a decision that a AAA member should be served by an ACME driver, within the limits specified by an AAA-ACME agreement, even though the member and the driver are not acquainted before service is needed and sharing no more information than is held on the plastic membership card.[1]

Prior approaches to trust establishment are based on two assumptions which are too strong for many situations—that trust between negotiating entities requires that each know the identity of the other, and that, in a hierarchy of certifying authorities, they have a common ancestor (Figure 1, left portion). Our contribution is simply to relax these assumptions; the technical content and the business values expressed in the rest of this report follow from this and from applying well-known techniques for reliably establishing the mutual trust two entities need to strike a deal, as suggested by Figure 1, right portion.

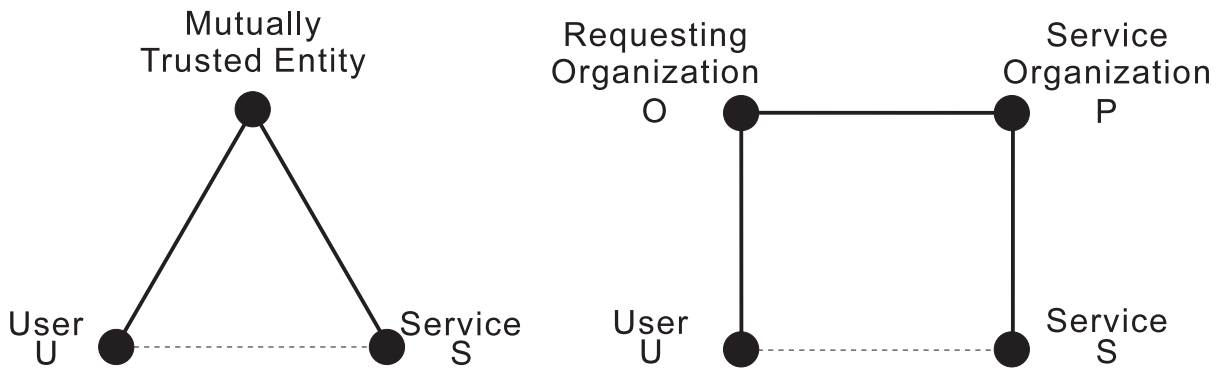

**Figure 1. Issuing certificates to establish trust.:** In the left figure, representing prior practice, the user **U** and the service **S** each gets a certificate that it is associated with a particular public key from a common, mutually trusted certificate authority. In the right figure, representing the new protocol, the user **U** gets from his organization **O** a certificate binding his public key to enrollment in that organization; the organization's agent and the agent of a service organization **P** are sufficiently acquainted to trust each other to make a service agreement, and the latter communicates this agreement to a server **S** that trusts such information from him. In each case, this is sufficient information for **S** to safely grant resource access to **U** when **U**'s request is accompanied by the certificate obtained. (Note: in each case, the trust establishment arcs might include intermediate certificate agents, as is well documented in the literature.)

---

[1] The usual form of the AAA plastic card in fact communicates its holder's name, which is in fact information not needed to consummate the transaction. It is included on the card as a convenience, e.g., to help manage lost cards. As far as we know, consumers are content that this case of extraneous information is not a substantial risk to privacy.

## Processes and Protocols

Figure 2 suggests pertinent network properties, viz., current Web servers which are mostly information delivery machines but among which state-changing in servers is an increasing component for e-business application and future embedded servers which are controlled solely by Internet-delivered commands.

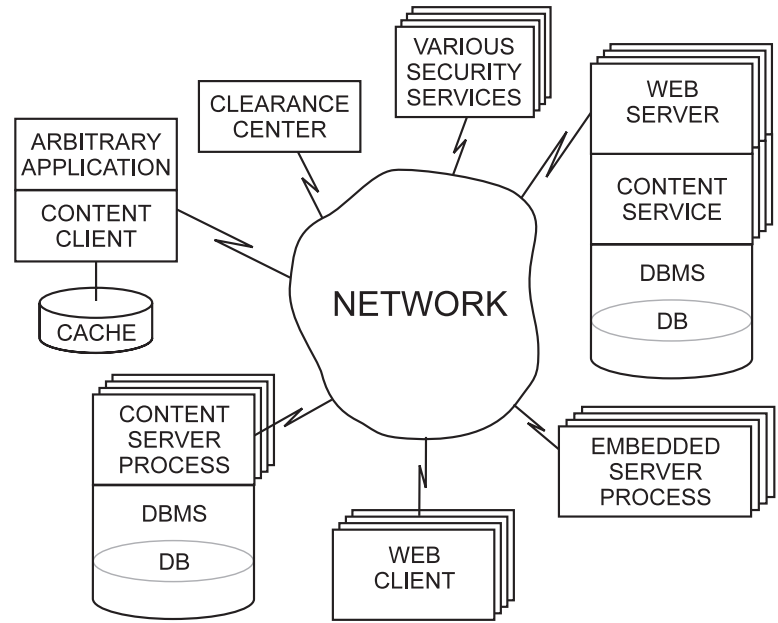

**Figure 2. Processes transacting via digital communications:** (clockwise from the upper right) web servers drawing on multimedia data collections; modest servers embedded in machines and appliances; web clients and whatever succeeds them as the Internet evolves; powerful content servers which might draw on the same databases as the Web servers; web clients and mobile clients, including untethered communicators; legacy applications which dominate today's manu-facturing and business production world and which are candidates for extensions to draw on digital content or drive remote processes automatically; and finally clearance centers, certificate services, and other network security services which collectively administer authentication and authorization.

### *Definitions and Scope*

When we consider authorization management (aka access control), it is convenient to classify user-to-resource relationships by labels.

*Insiders*   are users usually known personally to service enterprise management and/or computing service administrators.  Many of these users make database updates as essential parts of their jobs.  Authorization is managed by logon controls and decades-old access control services, e.g., [IBM RACF], [Saltzer 1975].

*Outsiders*   are peripheral users of the same resources as insiders, frequently querying and reading the same databases, but making database updates only as a side effect of read transactions, e.g., to create service logs and billing records.  Today their authorization is often managed by ad hoc methods, such as IP-address sensing [Lynch 1998], which tend to be regarded as no better than marginally satisfactory.

*Strangers*   are would-be users who might not be known to the services they access prior to the service requests.

*Incognitos* are strangers who choose to hide their identities and/or attributes as thoroughly as is consistent with obtaining requested services.

The current report addresses authorization for strangers and incognitos, and does not disturb procedures for outsiders and insiders. The server hosts envisioned range from very large machines delivering hundreds of services and millions of data objects to tiny computers embedded in appliances and having, as their only external interfaces, network connectivity through the appliance in ways not apparent to the casual observer, e.g., via the household electric power distribution with another household computer providing connectivity to the Internet. Any kind of server which can be addressed over an Internet connection might need the functionality we present.

The number of users authorized to exploit any particular resource ranges from a single individual to many hundreds of thousands of individuals. The authorization policies are often such that the sets of authorized individuals change too rapidly and too frequently for human administration of individual authorizations to be economical, or sometimes even feasible. Similarly, the resource pools might have rapidly changing memberships, e.g., in the case of admission to sporting events and in the delivery of exhaustible goods. The policies sometimes have complex rules which depend on end user attributes and service request circumstances, and these policies might intrinsically require distributed administration—partly by consumer administrators, partly by service consortium administrators, and partly by individual service offerers. <u>Automatic administration of encoded policies is often desirable and sometimes essential.</u>

In what follows, *unregistered* should be construed as "not registered by any administrator of the service in question". A *privileged* service, information, or good is a resource to which some requesters are granted access and others are not, with the decision often according to some published policy. The automatic mechanism described is based on computers holding and using data objects called *tickets* and *enrollments*, and on recognition that any computing service agreement can be represented by a mapping from enrollments to tickets.

A *ticket* is a token representing permission to use any member of a set of data and/or processing resources, just as in the non-digital world a bus ticket might permit use of any bus in New York. A ticket becomes useful when some user presents it or causes it to be presented to some server, and then only if the ticket is valid for the service being requested. Ticket validity may depend on *ticket modifiers*, additional conditions which are to be checked immediately before service delivery, e.g., the ticket may be valid only at certain times of the day [Gladney 1993]. Examples of tickets with associated modifiers are "printer use until May 19" and "one delivery of pizza".

An *enrollment* is a token representing a statement like, "Charles is an ACME employee in the purchasing department", i.e., a triple of a user identification, an organization identification, and a group identification within the organization. Other examples of enrollments are, "John has top-secret Navy clearance" and "John is granted e-cash worth $1234.56". In the latter case, the distinction between an enrollment and a ticket is fuzzy. An enrollment becomes useful only when it is held by a user and the user can sufficiently demonstrate that she is entitled to hold it. A *service agreement* or contract is typically a set of statements like, "ACME engineers are allowed to read IBM design documents for modems", i.e., a mapping from enrollments to tickets.

*Access control* and *authorization* indicate the same management activity—the former suggesting limitation of access otherwise permitted and the latter suggesting granting of access otherwise denied.

In access control, a *principal* is a surrogate for a service requester. In a rigorous treatment, the distinctions between *principal*, *user*, *process acting for the user*, and *user acting for some organization* would be very carefully made and followed. We ignore the distinction between an organization and its agents, whenever doing so does not lead to misunderstanding, just as we ignore the distinction between a human user and a process acting as surrogate for that user. As is conventional, we use the terms informally and assume that bindings are as carefully managed as is needed for safety; [Lampson 1992] shows this to be well founded.

We will repeatedly assume *reliable communication(s)*, whose transmissions are trustworthy both as to authenticity (content not inappropriately altered) and provenance (source claimed is the actual source) with a confidence level appropriate to the circumstances. This can be achieved in any of several well-known methods. More generally, *reliable* and *reliably* convey that the inference at hand is safe because it is protected by security measures which are implied by the context.

What [Kohl 1997] and we call a *clearance center* is approximately what [Seamons 1998] calls a *server security agent (SSA)*.

In applications of asymmetric encryption or public key encryption, we use $K_X$ or $k_x$ to denote the public key of **X** resp. **x**, and $J_X$ or $j_x$ to denote the corresponding private keys. For specificity in describing protocols, signing will be illustrated by the RSA scheme, in which the signature is an encryption under the private key of the signer. However any of several well-known means of signing may be used.

What we describe can be used together with widely known authentication ([Birrell 1986], [Lampson 1992], [Steiner 1988]) and certification services ([Verisign 1999]). We do not intend to reject these, but instead to emphasize applications that do not need them.

### *Processes and their Roles*

A server **S** manages a resource **R** permitted to requesters identified in an access control list which maps each user or group of users to a set of privileges on **R**. In addition to this conventional usage, the access control list maps each ticket to privileges on **R**. If the server happens to be a low-level machine without an administrative interface for human beings, this access list is reliably maintained by a connectable computer $A_S$ with security measures similar to those needed to load **S** with programs and other data.

Whether or not the loading computer $A_S$ (Figure 3 on page 6) also holds the server **S**, its human operator $A_S$ will have previously acquired the identity and public key of a clearance center **C** from its administrator $A_P$, who acts as an agent for some producer organization **P**. In many offices this will be easily done by human communication because the service operator and the administrator of the clearance center will be personally acquainted. I.e., it will be a safe "out of band" communication. The technical requirement is that this transmission should be secure (identities of recipient/source and message content all authentic). Also for administrators acting for a requesting organization and a producer organization, the parties will be personally acquainted, so that public encryption keys can easily be reliably shared.

$A_P$ and $A_S$ will also agree on the meaning of each ticket needed.

A clearance center **C** holds a map of enrollments to tickets for each server **S** it supports and a map of enrollments to enrollments for each consumer organization **O** it supports. It further holds a list of principals from which it will accept such information and any additional information needed to support extensions of its authorization services; each principal $A_P$ is the agent or administrator for a producer organization **P**. For the principal $A_P$, the identification is a unique name and its public key $K_P$ in an asymmetric key encryption scheme; this information is reliably acquired by well-known means. It is essential that $A_S$, acting for the server **S** which it loads with access control lists and other information, trusts **C** to issue its tickets; for this reason and because users and the requesting organization **O** needs to trust the intentions and actions of **C**, it must be roughly as trustworthy as a conventional authentication server or digital certificate service. This is easily done with a small computer under good physical control, e.g., under lock and key. Such a computer can be cheap enough so that clearance center is the only service this computer gives. For good availability, **C** and **S** must be reliably network accessible, but $A_S$, $A_O$, and $A_P$ need not be.

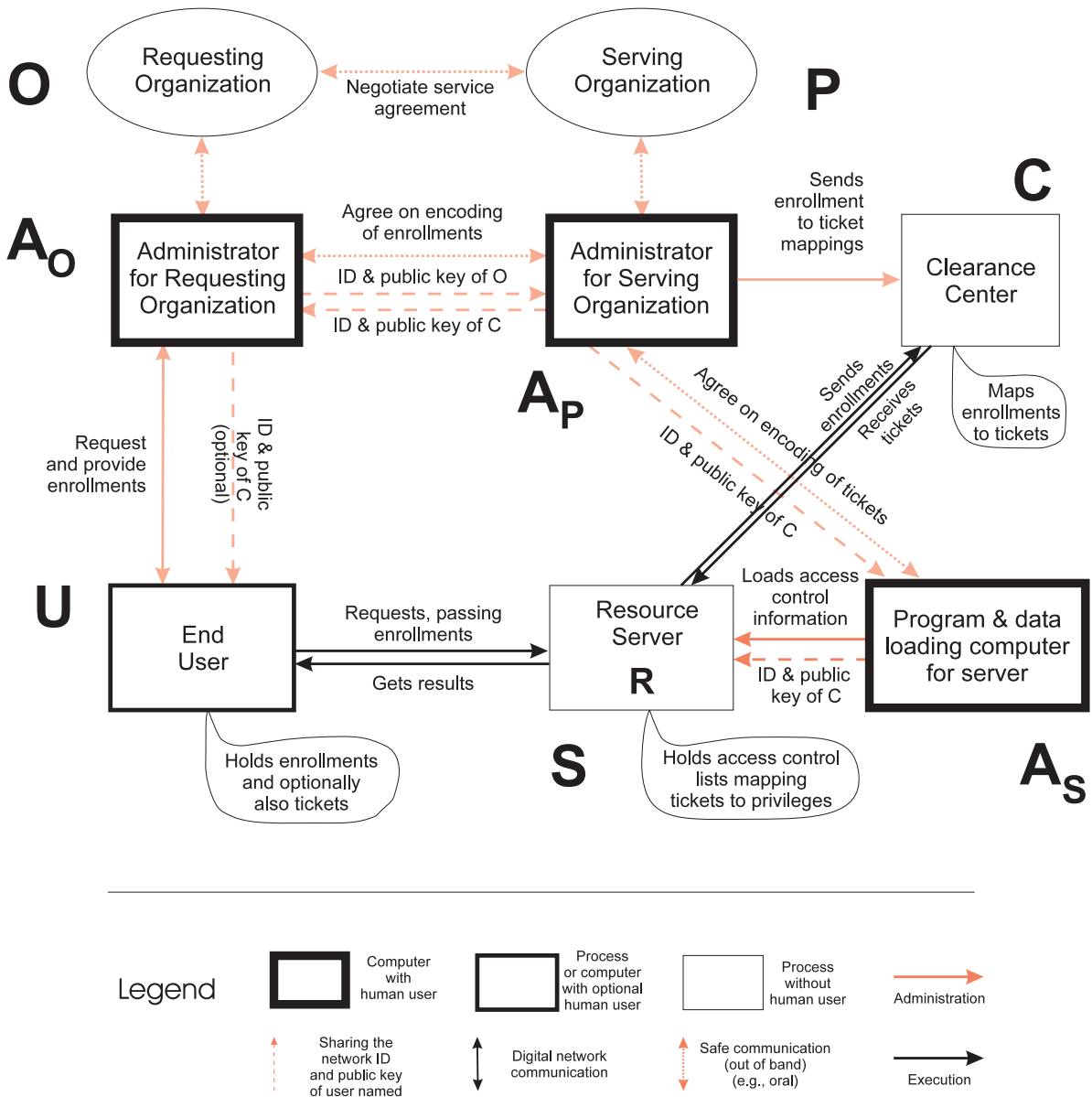

**Figure 3. Machines and messages for our preferred protocol:** as suggested in the legend, gray arrows depict administrative initiation communications; black arrows depict run-time execution communications.

Typically **C** represents many servers **S**, e.g., those for an office, the tenants of a building, or a company. **C** and the set of servers it supports use the same ticket value for resources deemed equivalent in all the servers (e.g., "print on a laser printer"); here *equivalent* encompasses only authorization. The representation of each ticket is simply a token unique among all tickets in the domain of **C**, and is typically associated with a character string external version for human comprehension. Other values, such as enrollments, are represented similarly, i.e., as a token and associated print string. The encoding of tickets for **C** and its cluster of servers might, without damage, conflict with that of some other clearance center **C'** and its cluster of servers, provided only that no server uses both **C** and **C'**.

An administrator $A_O$ of a consuming organization **O** defines and holds enrollments, which map users into sets of members equivalent for access-granting purposes, and a map of enrollments to enrollments which tabulates implications such as, "an ACME employee in the purchasing department is an ACME employee in the administrative division" and "an ACME employee in the administrative division is an ACME employee". The encoding of enrollments for **O** and its users might, without damage, overlap that of some other organization **O'** and its users. Enrollments are disambiguated by originator labels not explicitly shown in what follows.

Each end user **U** stores enrollments requested and received from $A_O$. Each enrollment is optionally armed or associated with an expiration timestamp which the user cannot modify. $A_O$ also gives **U** a cleartext version of this timestamp, which **U** can employ to refresh the enrollment set automatically. Optionally, **U** also holds expiration-time-stamped tickets which can be refreshed from the appropriate clearance center **C**.

### *Informal Description of a Protocol and Variations*

Whenever a service agreement is executed between a consumer organization **O** and a producer organization **P**, some agent[2] $A_O$ communicates to some agent $A_P$ the pertinent enrollments and supporting enrollment to enrollment mappings. The principal[2] $A_P$ acting for **P** updates **C** with enrollment to ticket mappings which reflect the service agreement and which further convey any agreed limitations on the use of each ticket, such as expiration date and so on. $A_P$ also stores **O**'s enrollment to enrollment mapping and public key $K_O$ and communicates the name(s) and public key(s) of the clearance center(s) **C** to $A_O$.

**O** makes available to each enrolled user **U** a certificate conveying the user membership class (aka enrollment) within the organization; this certificate is signed on behalf of **O**, i.e., with **O**'s signature. This certificate contains not only the enrollment, but also the identifier and public key of **U** under the signature of **O**. The form in which enrollment sets are delivered is either **([U E]$^O$)$^U$** or **({k$_U$ E}$^O$)$^U$**, which are explained below.[3] Whenever **U** suspects that her certified enrollments are not up-to-date, $A_O$ will respond to her request for a refresh. When **U** needs access to a protected resource, part of what she prepares for her request is the enrollment certificate signed by **O**. This she could send to **C** directly in order to obtain the ticket(s) it authorizes or, alternatively, include in the message requesting the service.

---

[2] An *administrator* or *agent* $A_O$ of some organization **O** is a principal that can sign for **O**. Since this notion of agency is precise and many-to-one, we can without ambiguity replace "the agent $A_X$ acting for **X**" with "**X**", and do so hereafter. This is an application of "speaks for" [Lampson 1992].

[3] We use the following notation for messages in the inter-machine protocol:

**e**        is an enrollment and **E** is a set of enrollments.
**t**        is a ticket and **T** is a set of tickets.
**Brackets []**        indicate that the contents are signed by, or on behalf of, the principal indicated by the superscript. E.g., **[x]$^U$** is **x** under **U**'s signature, i.e., **[x]$^Y$ ≡ x$^{J_Y}$**.
**Braces {}**        indicate that the contents are signed by, or on behalf of, the principal indicated by the superscript, using a ephemeral key pair (see the text) rather than a public/private key pair. E.g., **{x}$^U$** is **x** signed by the private part of an ephemeral key pair $k_U/j_U$; i.e., **{x}$^Y$ ≡ x$^{j_Y}$**.
**Parentheses ()**        indicate that the contents are encrypted to be confidential to the principal indicated by the superscript. E.g., **(x)$^U$** is **x** encoded so that only **U** holds the decryption key; i.e., **(x)$^Y$ ≡ x$^{K_Y}$**. (If there is no superscript, parentheses simply indicate grouping.)

Whenever **C** receives this, it enables **C** to verify that the request indeed comes from **U** and that the enclosed enrollment is indeed for **U**. If and only if these conditions are satisfied, **C** returns a **C**-signed certificate–the ticket **t** authorized by this enrollment, encrypting it, its usage limitations (e.g., duration) and the identity of **U**.

In a less-favored protocol variant, tickets are held by users for multiple requests. At the discretion of **P**, any such ticket may include an expiration time in order to permit unilateral ticket revocation. A well-chosen revocation time will typically be long compared to the intervals between expected mutiple appropriate uses of a ticket and short compared to the longest expected validity of the ticket, e.g., between an hour and a few days for human service applications.

As part of determining whether **U** is entitled, **C** uses its enrollment to enrollment map to determine whether **U**'s enrollment is part of a larger enrollment for which the ticket is authorized.

When **U** requests the resource, she sends to **S** the certificate, doing so under her private key $J_U$. By decrypting this certificate using first **U**'s public key $K_U$ and then **C**'s public key $K_C$, **S** can reliably conclude that **t** came from **C** for use by **U**, and can safely provide **U** any service protected by **t**.

This protocol lacks a desirable feature, privacy for **U** in the sense that neither **C** nor **S** can readily determine the identity of **U** or its other characteristics not pertinent to the service request. Some measure of privacy is provided by a signing such as that which follows in a slightly altered version of the prior protocol. (Privacy for requesters is new neither in the conventional world, where it occurs effortlessly in many transactions, including telephone calls in which the "caller-id" feature is not enabled, nor the digital world, where it is discussed by [Seamons 1998], inter alia.)

As part of providing enrollments to **U**, **O** also delivers a public/private key pair $k_U/j_U$ intended to be used only as a pseudonym for **U**, and hereafter called the *public/private* parts of **U**'s *ephemeral key*. Specifically, neither **O** nor **U** will reveal to anyone else that $k_U$ is associated with **U**. (This does not in itself prevent a server from ferreting out the identity of its user **U**, as **C** or **S** seeing $k_U$ in a message from **U** and needing to respond, might be able to get the network address of **U** to determine **U**'s identity. There are well-known ways of preventing this, but we doubt that they will be employed even by cautious users. However, nothing in our mechanism helps a malicious server to identify users.) Then the changed messages and interpretations are as before, except that any message containing $k_U$ is encrypted under the public key of its recipient (to prevent $k_U$ from being improperly substituted).

The end user can be further protected by limiting information accessible to either **S** or **C** to no more than is needed for the task at hand. Instead of passing all its enrollment certificates to **C** or all authorized tickets to **S**, **U** can pass just those tokens expected to be needed. How to implement this precaution is described in [Seamons 1998], which shows the more elaborate protocol required. A user might reasonably be further concerned that a malicious **C** could puzzle out her identity or a great deal about her by logging her requests and enrollments. This can be hindered by frequent changes of the ephemeral key pair $k_U/j_U$, doing this whenever **U** refreshes her enrollments by contacting $A_O$. The paranoid user could do this between every pair of requests.

The protocols above are suitable for information delivery, but not for requests which cause a state change or side effect at the server, because they are vulnerable to replay attacks on the **U** → **S** message. (For information delivery it is safe only if **S** responds using an address of **U** delivered reliably within the request and and encrypting the response under $K^U$; if this is done, the replay attacker will not receive the response.) Replay attacks are hindered by including another private token τ, described below.

## *Core of a Preferred Protocol*

We strive to satisfy the most stringent possible security objective, viz., that each process participant should communicate to other participants precisely and reliably whatever information that participant needs to accomplish what is requested, and no more information than this.[4] Further, network participants not called on within the transaction at hand should receive no information that would permit any of several well known misbehaviors.

Part of what is needed is to ensure the confidentiality of each message portion by encrypting it under a public key known only to the intended recipient—or set of recipients if substitutible network entities exist. A second part is to ensure correct content and provenance of each sensitive message portion by signing it with the private key of its originator, so that the receiver can be confident that it is not fraudulent, doing this for responses as well as for requests to ensure against "man in the middle" attacks.[5]

Our protocol (Figure 3 on page 6) is immune to a replay attack, uses no more messages than absolutely required, allows good privacy for **U** (e.g., acting as a customer) vis-a-vis **S** (e.g., acting as a merchant), and facilitates more powerful ticket modifiers than those already mentioned. This protocol does not require **U** to hold tickets, but simply passes encrypted enrollments to **S**, which passes them to a clearance center **C** for conversion to tickets.

Initialization and service delivery transactions may be accomplished in any order, except that an otherwise authorized service may fail if some machine does not yet hold the required enrollment(s) or ticket(s). The initialization actions are: loading enrollments into clients, doing so either as batch operations or interactively; loading ticket-to-resource mappings into access control lists in servers; and negotiating enrollment to ticket mappings and storing them in clearance centers. Safe protocols for these initializations are obvious and are therefore not be articulated.

The steps of a service transaction follow:

> **U→S: $(\{\tau\}^U$ R z $(k_U \{O [k_U E]^O\}^U)^C)^S$** The user asks the server for a resource **R** with parameters **z**, and includes his enrollments in a form interpretable by **C** but not by **S**. **S** uses the identity **R** of the resource requested to construct a set **T** of tickets **t** such that **t** would be sufficient to authorize the service **R**. $\tau$ contains **U**'s timestamp together with a nonce, and is signed with **U**'s private key. Everything is sent under **S**'s public key, to keep $\tau$ confidential.[6]

---

[4] The participants might gain more information from each other than is needed by other means than we discuss, such as by over-generous aspects of other parts of the communication protocol stack. Since considering such covert channels is outside our scope, our objective is in fact more modest, viz., not adding to the flow of unneeded information.

[5] The cautious reader will inspect the protocol for complete coverage. We are building a prototype, and will not be surprised if we find some attacks have been overlooked. In fact, for clarity, we have deliberately not included some protections, e.g., to inhibit repudiation of agreements entered.

[6] Since this message is somewhat difficult to read, we provide the following dissection.

$m_1 = [k_U E]^O$ conveys **U**'s public ephemeral key and enrollments, all signed by **O** to certify their validity.

$m_2 = \{O\ m_1\}^U$ conveys the identity **O** of the requesting organization and the prior enrollment information, so that the clearance center **C** can decrypt the enrollments, decide which enrollment to ticket map applies, and also verify that the consuming organization is in fact one that has negotiated a service agreement with **P**. This is signed with the ephemeral key of **U** so that the clearance center will be able to reliably check that it comes from the specific user to whom the enrollments were issued, but it does not include the request itself, to prevent the clearance center from tracking and correlating which services this user requests.

$m_3 = (k_U\ m_2)^c$ encrypts all this and also the public ephemeral key for **U** under the public key of the clearance center, so that **C** can interpret it, but **S** cannot. Thus the server **S** is inhibited from discovering not only the enrollments of **U**, but even what organization **O** this user is a member of.

$m_4 = R\ z\ m_3$ conveys **U**'s resource request **R** and associated parameters **z**, together with the prior confidential and signed

**S→C: (k$_U$ {O [k$_U$ E]$^O$}$^U$)$^C$**     The server **S** maintains a cache mapping nonces to timestamps received in some interval to tolerate clock discrepancies between **U** and **S**. If τ matches a cache entry or if the time stamp is outside the cache window, the request might be part of a replay attack and **S** does not honor it. Otherwise **S** caches τ and passes the signed enrollments to the clearance center. Neither **S** nor any user other than **C** can interpret this information, because it is encrypted under the public key of **C**.

**C: —**     The clearance center decrypts with its private key, getting the identification and shared ephemeral key of **U**. This enables the next level of decryption/unwrapping, which reliably communicates that the request indeed came from a user who claims his enrollments originate with **O**. Using the public key of **O**, which **C** was earlier given by **A$_P$** who has it from **A$_O$**, all by reliable communications, **C** next decrypts the innermost cipher, which communicates the enrollments and **O**'s certification that these were indeed granted to the claiming user. With this, **C** can trust that the claimant is authorized to these enrollments, and that this user is making the request.

If the clearance center finds that **E** authorizes some **t** in **T**:   **C→S: (t [t]$^P$)$^S$**,  the clearance center returns the ticket, together with its modifiers and signed with its own private key, to the server.

Otherwise:   **C→S: (failure code)$^S$**    the clearance center indicates that the enrollments do not authorize a key for the requested resource.

**S: —**     The server **S** decrypts to test first that the request indeed comes from whoever owns **k$_U$**, then that the tickets indeed were granted to this user by the clearance center **C**, and finally that **T** indeed contains the needed ticket **t**.

If all this is satisfied:   **S→U: (answer)$^U$**    the server does whatever was asked and returns whatever result is appropriate.

The messages delivered to the server **S** convey nothing about the user except what tickets he is allowed. The message to the clearance center **C** conveys nothing about the user apart from the enrollments he holds.

These are not the only command sequences that will work. For instance, the producer organization can pass the clearance center ID and public key to the requesting organization, which will share it with its users, enabling them to request tickets before sending messages to servers. A user might belong to several organizations which might entitle her to various grades of some resources. Such a user will want to be granted the best service level she is entitled to for each resource requested. This optimization is combinatorially complex in the number of network entities and enrollment tokens that need to be considered. [Ellison 1996] and [Herzberg 1999] touch on the problem; we have nothing to add to what they write about it.

---

|  |  | enrollment information. |
|---|---|---|
| m$_5$ = {τ}$^U$ | m$_4$ | includes a timestamped nonce so that **S** can exclude replay attacks. |
| m$_6$ = (m$_5$)$^S$ |  | encrypts all this so that only **S** can interpret what resource **U** is requesting, and so that no other machine can discover the values of the nonce and timestamp. |

The other messages are simpler, consisting mostly of pieces of the **U→S** request message.

*Refinements and Variations*

A variant would allow a previously unknown user **U** to establish an account and subsequently connect using that account and authorization to resources based on his tokens, which the server **S** saves as part of establishing the account.

The protocol specification in the prior subsection includes all the encryptions which might be required. Since public key encryption is moderately expensive, one or more of them might be omitted if the transaction participants deem their expense inappropriate for the risks at hand, or if the participants reliably know that other security means already preclude these risks. For example, a server **S** could establish a secure socket layer (SSL) link to each of its accepted clearance centers **C**, with this based on more efficient symmetric-key encryption; then the confidentiality encryptions **(message)**$^C$ from **S** to **C** and **(message')**$^S$ for the **C** to **S** reply would add nothing and could be omitted.

For clarity and simplicity, the discussion above has been in terms of a single enrollment, a single ticket, a single server, .... Obviously **U** could communicate to **C** all his enrollments and receive all tickets/limitations, and pass all these on to **R**, which would make accessible all the resources to which the tickets entitle **U**. [Seamons 1998] suggests this to preserve user privacy. **C** could act as the clearance center for several unrelated sets of servers. Such extensions are aggregations for performance and convenience which are a normal matter of program design, and are not further discussed in this report, beyond noting that they may require additions to messages and tables to distinguish between similar entitites.

Our descriptions have assumed separated server, clearance center, user and contracting organization machines. These processes which can be housed wherever is most convenient, provided only that they can be found when needed and that their advertised identities are trustworthy.

The clearance center **C** will be contacted by many users and might support many servers, i.e., its failure might impact service to many people. We can readily replicate it. Since **C** gets updates from very few agents $A_P$ and since the timing of updates need not be synchronized among replicas, this will be easily managed. If **C** replicas share **C**'s public/private key pair, the protocols described are suitable as is. For that illustrated by Figure 3 on page 6, the only change needed is that each server **S** should be given several clearance center addresses and should switch whenever its current choice fails.

The consumer and producer organizations **O** and **P** might be the same organization, and their agents $A_O$ and $A_P$ might be the same principal. Even when **O** and **P** are distinct, they may agree on a single agent $A_O \equiv A_P$ to act as a trusted broker. Doing so will reduce the human administration needed somewhat and also enable the use of tickets in access lists within a single organization, which might be useful because sometimes tickets are more convenient than user/group entries in access control lists.

Our mechanism can be used with smart card technology. We would load the enrollments and tickets onto a smart card. Whenever **U** wants a service, he would plug the smart card into the appropriate server **S**. **S** would route the enrollments to **C** for resolution into tickets, and return these to **U**. After that **S** would deliver the service if it received the correct ticket(s) from **C**. As shown, the transmissions of enrollments via **S** would be encrypted so that **S** could not interfere with them or eavesdrop, helping to preserve **U**'s privacy. All the privilege checking would be hidden from the human user, **U**, except for whatever extra time is needed for checking and any "you are not authorized to do that" messages it causes.

The preferred embodiment attempts "complete" security, i.e., that every identified threat is countered by an effective fence. This, of course, is not without cost; specifically, public key encryption is expensive in machine cycles and elapsed time. Practical implementations may choose to forego some encryptions for confidentiality when the potential losses consequent upon eavesdropping are small. Obviously, if it is reliably known that the communication channels are encrypted lower in the protocol stack, it is not necessary to encrypt for confidentiality at the protocol level we have described.

Similarly, there may be circumstances in which signing is not needed to prevent masquerading. For example, if a session between some client and some particular server involves many request-response cycles, the nonce included in the first **U** to **S** request could be repeated in subsequent requests as demonstration that these come from the same entity that signed the first request. (This is safe if it is within encrypted requests. If these subsequent requests are not encrypted, it is still a significant nuisance barrier to unsophisticated attackers.)

### *Constraints on Enrollments, Tickets, and Services*

An enrollment can have embedded or associated *enrollment modifiers* which convey limitations on its use, i.e., modifiers deny services which might otherwise be available. The most obvious limitation is an expiration date, but many other limits might occur, such as dollar value limits or any other condition that designers of servers and clearance centers can arrange to be tested.

Similarly tickets can have associated modifiers which convey limitations on their associated service deliveries. Further, servers could include further modifiers as part of access control lists. The ranges of ticket and server modifiers are similar to those of enrollment modifiers.

The general form of a modifier is a predicate function of constants and variables which can be represented with a procedural or non-procedural programming language and can be evaluated in whatever machine is appropriate for the tests at hand. For an enrollment modifier, this would normally be the clearance center **C**. For ticket or server modifiers, it would normally be the server machine. More elaborate modifiers are possible, requiring comparison between the internal state of **P** and that of **S**. [Seamons 1998] suggests language constraints to facilitate client-side evaluation of authorization rules, and [Bacon 1998] treats related aspects.

Modifiers are evaluated partially before tickets are granted and completely as part of **S**'s finally decision whether or not to grant **U**'s request. The complete modifier is the combination of modifiers from the enrollment which yielded the ticket, from ticket modifiers held by **C**, and of further modifiers held by **S**. For example, **O** could grant **e** to **U**, "only for calendar 1999". **C**'s mapping **e**→**t** could specify, "terminating at end-September 1999", and **S** could specify, "**t** is good only in prime shift" and further, the access list for **R**, which is allowed via **t**, could specify, "only with background priority". The effect of these particular modifiers would be to allow service only during normal business hours in January through September 1999, and to delay service until all higher priority service was complete.

A ticket might require a mini-payment [Abad Piero 1998] or a banking transaction as a condition of release. The clearance center can engage in banking transactions as part of the clearance, with the enrollment including account information, and there could be information to the user as to what his account will be charged, allowing him to approve or reject this before the transaction is committed. Such procedures are not novel and are mentioned here only to point out that they are compatible with broader mechanisms.

*Debiting modifiers* are of particular interest because they emulate many ordinary-world practices, such as using postage stamps. When an enrollment or ticket is granted for a depletable service or good, grantors will typically want to limit the service quantitatively; this is most obvious and common in transactions involving money, but might be wanted for any countable or measurable resource. Possible are four flavors of debiting modifiers: each holding either an integer or floating point quantity, and each optionally requiring end user affirmation.

Our preferred protocol for debiting transactions is an addition to the protocol described above. Whenever a clearance center **C** hands to a server **S** a ticket **t** whose authorizing enrollment **e** has a debiting modifier **m(v)**, **C** stores for the duration of the transaction a correlator for **e**, **t**, and **m(v)**. (This is necessary because there may be more than one **e,t** pair that makes the requested resource **R** accessible. **v** is the current quantity held by the modifier.) If the modifier requires end user confirmation of the transaction, it asks to **U** approve (just as it might do if the quantity were a monetary price), together with character strings describing units of measure, debit amount, transaction description, and other information for a possible human user. This information will have been carried as part of or otherwise associated with the debiting ticket chosen by **S** as the authorization to deliver **R**. If and only if **U** confirms, **S** will identify to **C** the ticket **t** it selected, together with the debit amount. **C** will appropriately debit **m(v)** for the associated **e** and signal **S** to proceed with service.

# Further Work Needed

## *Performance*

The elapsed time costs of transaction consummation is likely to be primarily incurred in executing inter-machine communications, with execution-time costs being more important to users than the costs of less frequent administrative initiations. Figure 3 on page 6 illustrates that the inevitable cost of transations is a single request/result message pair, and that our protocol adds to this one more request/result message pair. Improvement over this seems unlikely.

The user who wishes best-possible hiding of her identity and attributes must additionally renew her obscured enrollment certificate for each and every transaction, and must pay the price of an additional message pair.

Some users will be members of many organizations. Some servers will contribute to more than one server organization. We have not yet considered the consequent complexities, but believe that the cost of giving each user the most privilege she is entitled to is a combinatorially complex choice from mappings held in the clearance center, with ramifications elsewhere in the network. The extent to which this might lead to user-visible delays is among the topics yet to be considered.

## *Consumer Privacy*

Consumer privacy is currently much in the news and figures prominently in professional conversations. For example, in a recent executive education session [Stanford 1999], several speakers called for computer industry help with this and closely related challenges. One of these speakers was Larry Irving, Assistant Secretary of Commerce for Communications and informtion Administration (NTIA); his discussion, consistent with that of two other speakers, emphasized that current transaction procedures and practice often required consumers to reveal far more personal information than was reasonably needed for provider safety in the transaction.

For a large number of commercial transactions, the technical objectives[7] seem to be:

1. that each machine involved in a transaction should have access to no more information about the principals of other machines than it needs to enable that part of the transaction it must accomplish for the entire transaction to be consummated;
2. that the transaction should be divided among machines so that no machine can "put it all together" (division of responsibilities); and
3. that relatively risky information should be exposed only in machines that are relatively trustworthy.[8]

We believe, but do not in this paper demonstrate, that the protocol above, or straightforward engineering extensions of this protocol, satisfies these objectives and is optimal in the sense that no other protocol can provide transaction management with better privacy protection. For the moment we leave this as a conjecture to be addressed in future work.

---

[7] To keep this articulation concise, it does not include careful definitions of the risks and trust levels alluded to and also adopts an informal style. What is written here informally can be articulated in a careful manner that is both precise and comprehensible to a specialist. With additional effort such an articulation can reach outside professional circles, to people who might not intuitively understand why high precision in this is not only desirable, but even necessary.

[8] This last objective is what motivates the clearance center use, because the operator of a properly limited and managed clearance center cannot by herself defraud anyone. Further, the operator of a clearance center depends on her reputation for integrity if she is not to go out of business.

For each request to include no more user attribute information than the server requires, the end user's agent must know what end user information is needed. We have not addressed how it can get to know this; surely it requires more communication than Figure 3 on page 6 suggests. We intend to address this within the context of prototype building.

### *Sharable and Private Tickets*

The new protocol has been designed to mimic real-world (non digital) behavior as much as possible. This mimicry might not always be what people want, e.g., when tickets are handed to consumers, because without further constraint such tickets can be copied or transferred to other consumers.[9]

Of course for some applications this might be precisely the behavior wanted, or there might be little value to constraining tickets to effectiveness only to the user to whom a clearance center made them available. For instance, if the resource authorized is read access to a document, any recipient could pass a copy to some third party; there would be no point in denying that such a read ticket could not itself be passed to the third party. In contrast, sharing of an update-authorizing ticket certainly would change the data integrity commitments of the transacting parties.

Thus customers might want implementations which automatically constrain update tickets differently from read-only tickets. Unfortunately we cannot provide this because it depends on the behavior of the resource server, which is outside the scope of the protection mechanism, and further because it encounters the program decidability problem.

What we can do is provide the service offering enterprise with the ability to distinguish between read-only tickets and update tickets and further to enable choosing the ticket kind as part of specifying a service agreement. One way to do this is to assign a common asymmetric key pair $K_s/J_s$ to all servers **s** that accept the same ticket(s). When a consumer **U** requests of a clearance center **C** some update ticket **t**, this is returned together with the public ephemeral key $k_U$ of **U** all encrypted under $K_s$. This would be accepted by **s** only if it is further signed by **U** with the private part $j_U$ of its ephemeral key.

Note that this protocol does not prevent **U** from getting keys for some other user, but rather ensures that each set of keys delivered by **C** is usable only by some single user for which it is requested. I.e., a user **U** can request for some other user **U'**. Further, users can collude to defeat this intent by sharing private ephemeral keys, and the scheme further allows any receiving user to employ the key repeatedly. Although we believe such behavior can be thwarted by a bit more cleverness about assigning and using keys, we have not thought it through. Given that our preferred protocol avoids the problem, we are not currently incented to do so, and leave it to that reader who finds a sufficient motivation.

### *Scaling*

Figure 3 on page 6 and the protocol articulations have been presented as if each user **U** belonged to a single organization **O**, and as if each server **S** was related to a single serving organization **S**. In fact it is quite common for a human user to be enrolled in many organizations—a school, a church, a club, a city, ...—and possible for a resource server to be an offering of many service organizations. In general, every entity in Figure 3 on page 6 will be associated with a multiplicity of each other class of entity with which is related.

Within certificates, the ambiguities that such scale up would otherwise introduce are already handled by fact that each certifying entity signs the certificate with its network-unique identifier. If we extend this by including the network node identifier of its creator as part of every enrollment and every ticket, no ambiguities will occur. Of course the identifiers must be *Universal Unique Identifiers (UUIDs)* [Gladney 1998b].

---

[9] This problem does not arise in our preferred protocol, because the tickets are not passed to be held by consumers.

For a requester that belongs to many organizations, the step of mapping from enrollments to tickets and the number of candidate tickets for the request of the moment could become quite large. Such cases might be optimized by protocol additions such as having servers **S** respond to inquiries such as, "What ticket is needed to deliver resource **R**?" and having clearance centers **C** respond to questions such as, "For members of organizations $O_1$, $O_2$, ..., $O_n$, what enrollments would permit usage of ticket **t**?"

### *Implementation and Possible Realizations*

To help us understand which optimizations are worthwhile, and to design for minimal human administration, we are building a prototype implemented with the Java® language, Web browsers, and DB2 UDB databases. Figure 4 suggests the subset of Figure 2 on page 3 that we intend to include. For the server **S**, our authorization design [Gladney 1997] is upwards compatible from what is in the IBM Digital Library® product [IBM VisualInfo], which needs only easy extensions to accommodate tickets.

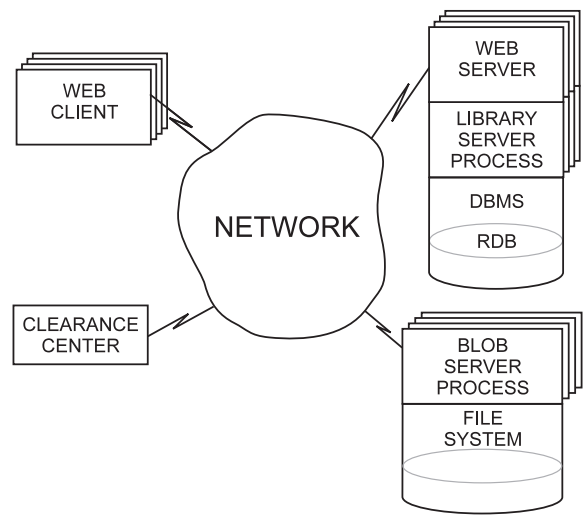

**Figure 4. Network layout for content management services using authorization for strangers:** illustrates a prototype which is in early stages of design and build to test and demonstrate how we can achieve large performance improvements over today's IBM *VisualInfo* product.

Our clearance center implementation will be based on the code of the clearance center that is part of the IBM Cryptolope™ offering. One extension we need to consider is the part of a rights management language needed to represent modifiers. (Representing enrollments and tickets is trivial, since they are only otherwise meaningless tokens accompanied by print strings.) For that we'll draw on AT&T work **[citation needed]**, IBM Haifa Research Laboratory work [Herzberg 1999], Xerox work **[citation needed]**, and similar efforts, to the extent possible consistent with an XML and RDF framework **[citation needed]** for the transport representation.

Of interest is use with SUN Microsystems's JINI™ network services. Attractive possibilities for pilot applications include:

**Personal servers:** home appliances and automobile amenities (such as a heating block for cold climates), with a limited number of known users authorized for remote control.

**Commercial services:** purchase payment with client computers buried in cash registers, and customer's smart cards carrying the enrollments and involved in confirmation of banking transactions; specialized printing, drafting, and photographic processors, to which pictures are sent while you drive to pick up the hard copy; ...

**Office services:** printers and projectors driven remotely; remote-control surveillance devices.

**Client machines:** hand-held or pocket wide-band communicators which call for information from a remote source to be delivered and also forwarded to

**Personal authorizations:** An attractive medium for enrollments and tickets is a smart card, which the user loads with enrollments by using it on any computer and having that computer contact the enrollment-issuing computers on his behalf, and then uses at some other time and some other computer to turn enrollments into tickets for the server set of immediate interest.

System security will be no better or worse than the underlying encryption and key management tools, which have well-known limitations [NRC 1999]. Clearance centers are similar to authentication servers and certificate authorities in their sensitivity and the protections needed for them.

The enrollment tokens held by individuals are short–less than 5000 bytes each. For applications which do not require much data to pass between **U** and **S**, the communication might be via human skin transmission and a handshake [Zimmerman 1999], with **U** implemented as a personal appliance computer, or by the human user touching the control knobs and switches of an appliance with an embedded computer, i.e., without any human action beyond use of well known appliance knobs.

## Discussion

Our permission decision process will be easily understood by many people because it mimics established common practices such as those associated with postage stamps, mass transit tokens, "rain checks", discount coupons, and so on. What's new is recognizing that yet another common human process can be automated because digital communication is becoming ubiquitous, and that this automation reduces human overhead so effectively that fine-grained control of very small resource pools is suddenly practical.

### *Prior Work*

Those who work with methods for establishing trust in distributed systems seem to have overlooked that enrollments are different from tickets and that privilege evaluation sometimes should exploit enrollment $\rightarrow$ ticket mapping. This may be because until 1998 no-one had directly addressed the problem of deciding whether or not to serve an anonymous or previously unknown user, except when the service is available to anyone or money changes hands.

What we have discussed is complementary to and can exploit tools for micro-payments [Abad Piero 1998] and safe electronic signatures [Herzberg 1998]. The implementation will exploit IBM's Cryptolope® technology [Kohl 1997], which up until now mostly helps manage secrets needed before a decision to deliver information or after such a decision. The current work helps manage the decision to release or withhold service, and administer the information needed as input for this decision.

Related work [Ellison 1996], done in the context of the IETF committee on the Simple Public Key Infrastructure (SPKI), carefully discusses trust and risks in the acquisition of identities and certificates from acquaintances. But it starts by assuming that the user's identity needs to be known to grant authorizations and never abandons that assumption. Instead it focuses on identifiers good enough to issue authorizations (Figure 5), and points out that producers can safely serve strangers if they are willing to consider authorizations to be bound to digital identifiers rather than to legal entities, i.e., by giving up the notion that digital identifiers are connected to individuals or enterprises.

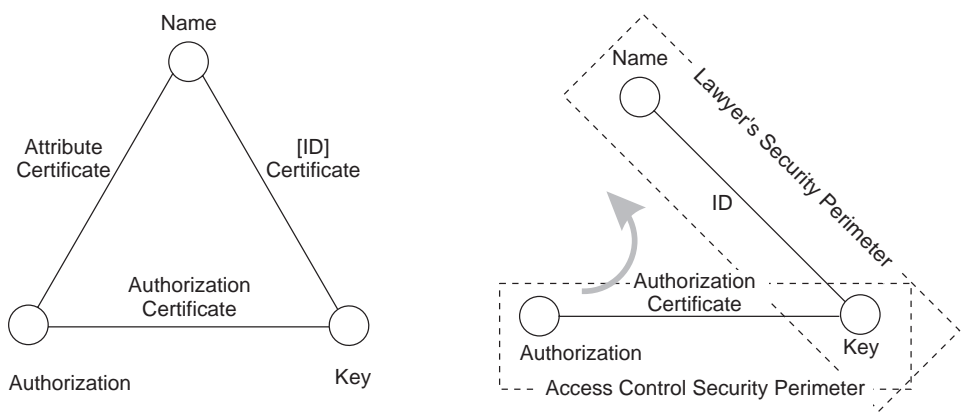

**Figure 5.  Authorization (ticket) handed to authenticated user** (from [Ellison 1998])

[Seamons 1998] focuses on specifications of credential acceptance policies (CAPs), with resolution of each CAP both on the server side and on the client side in such a way that a client can reveal only such user description portions as are needed to obtain some requested service. But in discussing what it calls "credential chains", it continues to assume that producer and consumer must share a common trusted authority, viz.,

"In order to authenticate a credential, such as a student ID, it is often necessary to have a supporting chain of other credentials.  A student ID must be signed with the private key of the university, so authentication requires knowing the university's public key.  A separate credential may be required to authenticate that public key.  For instance, this supporting credential could be signed by a certification authority and state that the university's public key belongs to the university.  (Several corporations provide such a certification service today.)"

[Seamons 1998]

This work is being continued in Transarc Corporation's Digital Credentials project:

"Digital credentials will enable clients to gain easy access to secure web services.  They are analogous to the paper credentials we possess today.  This project focuses on supporting services that do not issue their own digital credentials and that may have no prior relationship with their clients, but that need to know something about their clients nonetheless.  In our system, credentials that were issued for other purposes are accepted by the service.  This is analogous to the way we currently use a drivers license to prove identity when cashing a check."

<w3.transarc.com/afs/transarc.com/public/trg/projects/DigitalCredentials/>

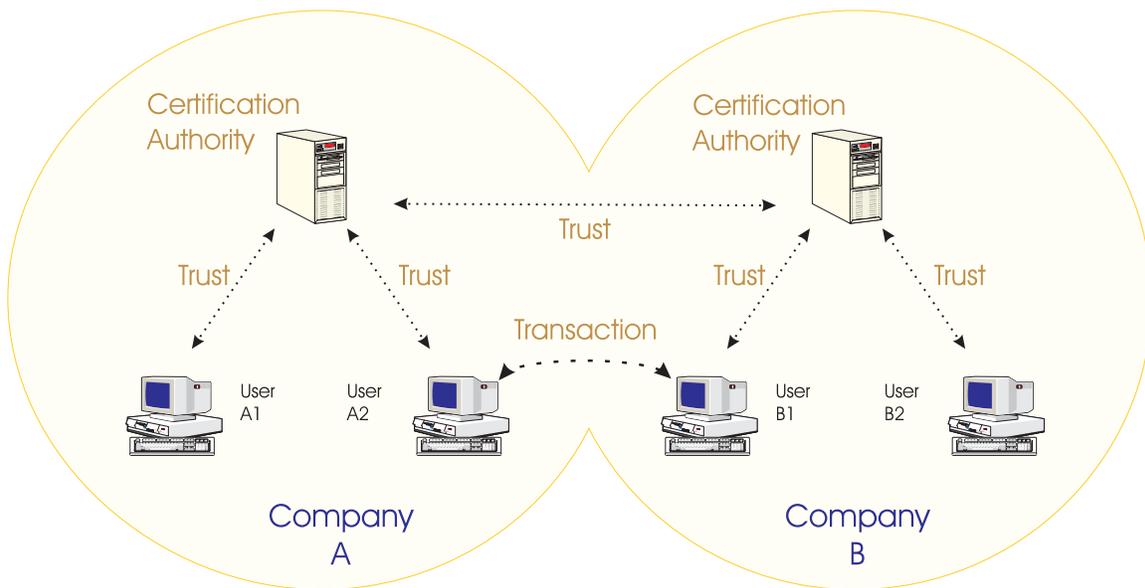

**Figure 6.  IBM S/390 team vision of inter-company trust model:**   other people are beginning to consider the kind of inter-system relationship we describe, without yet having considered the use of tickets instead of user roles.  This diagram was shown in a June 1999 IBM-internal meeting.

Independently and simultaneously with the current work, [Herzberg 1999] applied role-based access control ([Sandhu 1998]) to much the same problem, emphasizing an XML-based policy language to transmit certificate to role mappings.  As in the current proposal, this work relieves servers of having to recognize particular users by having them hold attributes describing authorized users.  [Herzberg 1999] differs from the current work by having one fewer level of indirection between the end user's identity or identity certificate and any resource permission found in a server's access control list; it puts role identifications (similar to our enrollments) into access control lists, rather than tickets.  This means that whatever administrator negotiates the meanings of roles with servers must also manage the assignments of roles to users.  This will become unwieldy for large and constantly changing pools of users and servers.  It will be quite unworkable when each user belongs to many requesting organizations, when many servers act as subcontractors for more than one serving organization and are related to more than one clearance center, and when requesting and serving organizations each deal with a multiplicity of serving resp. requesting organizations. We believe that stranger-friendly authorization services cannot have fewer indirections than suggested by the right hand portion of Figure 1 on page 2. I.e., for service networks that can grow to be very large and contain entities which each deal with a multiplicity of other entities, we believe our method to be superior.

Of course money and financial instruments can be represented by special kinds of tickets; ownership of a credit card represents a particular kind of enrollment. Other workers have anticipated these special cases, they are coming into use in e-commerce. Part of what is novel about our method is that the protocols and encodings described extend prior methods to any kind of ticket used by a small producer organization and any kinds of enrollments that might be used in acquiring a local service, e.g., access from one company into selected data resources of another company, or family membership to permit home appliance control.

### *Advantages Over Prior Work*

Our mechanism is a new digital service–a server contacted on behalf of previously unknown user can automatically decide whether to deliver privileged service, and do so consistently with policies specified in service agreements. It reduces essential administration enough so that very small organizations can offer or acquire service. Specific properties achieved are that:

1. Both active resources (e.g., your home furnace) and passive resources (e.g., information) can be controlled.
2. Realizations can be self-contained in the sense that no outside authentication service or certificate authority is required.
3. Servers can unilaterally add or remove resources, or change which tickets permit which resources. Such changes become effective immediately for ticket holders.
4. Clearance centers can unilaterally change mappings from enrollments to tickets. Such changes become effective immediately to users who present enrollments to get tickets.
5. Requesting organizations can unilaterally add or delete enrollment classes. Additions become effective for a user as soon as she gets the **O** enrollment (if she is permitted it) and $A_P$ includes them in the enrollment to ticket mapping. Deletions become effective as soon as $A_P$ removes the associated entries in its enrollment to ticket mapping.
6. Neither tickets nor enrollments can be transferred for use by unauthorized users.
7. Whenever desired, tickets and enrollments can be armed with expiration timestamps. This allows clearance centers to avoid the well-known problem that tickets are not revokable, but can still allow a ticket to be reused many times before it expires. This is safe because the only clock synchronization it depends on is that between **S** and **C** in the case of tickets, and that between $A_O$ and **C** in the case of enrollments.
8. Many different ticket modifiers, including types from prior work [Gladney 1993], can be used. They include tickets with limited service periods (time of day, start date to finish date, etc.), tickets for a limited number of uses, tickets which debit a value and become ineffective when this reaches zero, tickets with associated parameters which limit the service to particular instances of a set of similar resources (e.g., seats in a theatre), encryption keys, and so on.
9. Debiting ticket modifiers can track all kinds of service measures and can cover resources from any subset(s) of servers associated with a clearance center.

10. User identities are optionally hidden from servers and clearance centers, except that a statistical attack mounted from a clearance center might succeed against a user who does not change her ephemeral key often enough. Of course, any server might choose to deny resources to incognitos.
11. Various machine layouts and network protocols (order and content of messages) are feasible, depending on what is convenient for the circumstances at hand.

Although this is not unique, we point out that the human interfaces at $A_O$, $A_P$, and $A_S$ can be simple and intuitive, as can the $A_O$-to-$A_P$ and $A_P$-to-$A_S$ communications, because all these mimic brief human conversations.

### *Value to Research Libraries and Their Users*

Fortuitously, just as this report was nearing completion in early April, Judith Klavans sent the report of a Digital Library Federation workshop on access management [Arms 1999]. This gave the unusual and welcome opportunity to evaluate the current work against an unbiased requirements statement from representatives of key North American research libraries.

Each university enrolls each of its members from time to time (for undergraduate students, typically at the beginning of each academic year; for faculty and staff, typically whenever a signficant change of status occurs; etc.). As part of enrollment, each potential library user can be given certificates binding her to some group(s) which have library privileges; such bindings are not limited to library services, but may have other associations. Each enrollment is stored in a university membership registry which will be checked as part of logon authentication, or on a smart card, or somewhere else which permits token retrieval during an on-line session.

If other groups have claims on library services, as in Indiana, where statute grants every citizen the same university library privileges as students have, some other agency than the university can do similar enrollment. Later checking steps parallel their university counterparts in obvious ways.

The executive summary of [Arms 1999], which can be seen in its entirety on the Internet, poses questions and tabulates needed properties of an access management system. We reproduce below the core of every question and every requirement and answer resp. comment on these. The questions are:

>*How can members of a university that has subscribed to an electronic journal prove that they are authorized to access an article?* They can present enrollments or tickets as described in our protocols.

>*How is a system to confirm that the staff member, professor, or student is not someone else?* At the time a user enrolls, (s)he must demonstrate his/her identity as already required by existing enrollment procedures. At the time a library service is requested, (s)he does not need to prove identity, but only that (s)he holds an appropriate token.

>*Are there ways to screen out impostors?* Our protocol includes means to prevent improper sharing of enrollments.

>*How finely can information providers discriminate among potential users when making their materials available?* They can define as many enrollment classes as they wish. For each enrollment class, they and the service offerer must agree on a slate of privileges, which the service offerer must reflect in his loading of clearance centers.

*What criteria should universities ... use to determine who should have access to a database of published information ...?*  This is a policy question with which no automatic mechanism can help.

*What options do public libraries have ... to authorize the use of licensed materials to the general citizenry ...?*  How to address this is suggested in the third paragraph of this section.

*How can authors and other creators of information resources be protected from digital thievery?*  As suggested partly in [Gladney 1998] and further in *Safeguarding ...* articles it cites, no technical means can inhibit individual and private misappropriations.  Further, legal and constitutional constraints on search and seizure inhibit effective protection in most jurisdictions.  In contrast, wide-scale distribution of reproductions, called *piracy*, can be detected with cryptographically-based watermark schemes [Mintzer 1997].

*[Will] authors become "the deer in the headlights" of a vast traffic they cannot control?*  Authors' problems are principally inherent in their contractual relationships with their employers (the universities) and their publishers, i.e., beyond what technical means can help with.

*[How can] custodians ensure [accessibility] but that proprietary rights ... are protected ...?*  Custodians should mark on-line versions with watermarks to inhibit piracy.  [Craver 1998] and other articles in the same issue of Comm. ACM comprehensively summarize digital watermarking and its limitations.

*Should digital data be fitted with a digital lock that can be opened only by users with matching keys?*  This is a policy decision which will be taken by copyright owners or their agents.  Libraries and end users might be able to influence such decisions through market forces.

*[Compatibility] with constitutional and legislative [balancing] rights of ... creators and citizens ...?*  The U.S. Copyright Act of 1976, which defines "fair use", makes fair use a defence against a charge of infringement.  Some people argue that the U.S. Constitution intends to make forced disclosure of intellectual content a right of anyone who demands access for public good applications.  Any attempt to make the law over in this direction is sure to be vigorously opposed.  Part of the opposing argument is likely to be that such forcing would jeapordize the right to keep secrets–either private secrets or enterprise secrets such as proprietary information.

The requirements statements, slightly condensed, are:

**Simplicity**.  *The less complex a system is ... the more acceptable it is ...*  Our new mechanism permits implementations whose only new administrative actions are very easy steps within existing enrollment procedures for participants and straightforward recording of the essence of service agreements.  <u>Good implementations will require no extra human action when a service is requested and delivered</u>, except when the consumer wants the option of validating, e.g., for a transfer of funds.  We cannot imagine a simpler interface.

**Privacy**.  *... of users from detailed tracking and disclosure of use.*  We've shown how end user privacy can be achieved.  Of course, some producers will choose to deny service to anonymous users.

**Good faith**.  *... Users and providers would prefer ... reasonable barriers against abuse rather than ... inhibit use.*  Whether or not a barrier is reasonable is a judgment that we cannot make.  However our mechanism is neutral vis-a-vis the rules that providers and intermediaries choose to grant or deny service.

> **Trusted intermediaries**.  *Intermediaries play an essential role ... as parties trusted by both users and providers ....  System design must take the role of intermediaries into account*.  The essence of what we propose is to depend on intermediaries to act as brokers between producers and consumers.
>
> **Reasonable terms**.  *... not limit access to specific user groups ... but ... open to serving unlikely users whose curiosity ... may lead them in directions not predicted ....*  Whether or not any specific terms are reasonable is not for us to judge, beyond pointing out that nothing in our mechanism affects the terms that providers choose to require of their prospective users.

Briefly stated, these assessments in the domain of automatic authorization management suggest that our mechanism could help people achieve the objectives of the libraries in whose interests [Arms 1999] was written.  The help results from our having introduced one more level of indirection than is inherent in mechanisms considered by [Arms 1999] and [Lynch 1998].  No technology can address questions of policy or law in resolving opposing interests, so our work is not applicable to some of the questions above.  A more careful examination of our mechanism by the research libraries is sure to confirm this cursory assessment.  We welcome critical evaluations.

## *Business Value*

The business value of our mechanism results from its enabling access by many network users to many network-accessible resources without any individual user or the human manager of any resource needing to negotiate access, or even anticipate which resources resp. users they wish to use resp. provide.  An early application is protection of commercial and personal digital resources which individually are not inviting targets for misuse, i.e., for which elaborate or expensive security measures are not warranted.  For such resources, small percentages of breached policy are accepted as reasonable business risks, and this mechanism can achieve its objectives in commonly deployed environments without additional measures.  More generally, services envisioned include:

- controlling home/office appliances, e.g., printing to other people, fetching confidential correspondence;
- validating parking after goods purchase and getting other commercial services;
- generating a shipping order by an action of a purchasing agent of some other company;
- getting courtesy services, such as British AA services granted to AAA members;
- policing entrance to restricted facilities and managing access to theatrical performances; and
- delivering the encryption key of a super-distribution.

Because our method does not need outside authentication or certificate services, small-scale implementations are cost-effective.  Any would-be service offerer can implement an agreement with any acquiring organization, without assistance of third parties and without opening any security exposures not already inherent in his environment.

In addition to commercial and private application, we think the new approach will facilitate administration of various public safety and legal applications such as control of traffic signals and other police applications, authorization of financial and security auditors to needed information, legal discovery, draft legislation, and so on.  We believe it will also be valuable for very sensitive applications such as military intelligence if it is used together with other security techniques, but have not yet examined this conjecture carefully.

Human administration needed is minimal and can be implemented to be intuitive.  The only administration necessary on the user side is that some agent of each organization in which the user is enrolled (e.g., university, credit card provider, employer, club, ...) must supply her with a certificate of the enrollment class. The only administration necessary on the server side is that it and the agent of the producer organization must agree on the meanings of service tickets. Finally, the two agents must agree on a mapping of enrollments to tickets–the representation of a service agreement.

Today, network service offerers typically require each user to enroll with that service and to remember her password. This is both inconvenient and gives up some user privacy; while the service offerer may want the user information for commercial exploitation, users are not pleased with this. Our mechanism does nothing which helps anyone invade user privacy; of course it cannot close Internet or other loopholes that allow user identities to be discovered from messages they send.

Administrators can enhance or revoke consumer privileges quickly on the consumer organization side (to change what some user(s) can do), on the producer organization side (to effect service agreement changes), or at any server site (to add or withdraw particular services). Quick revocation is immensely valuable, because it makes possible rapid reaction to risks like fraud.

In summary, by reducing human administration to a minimum, we allow very small organizations to manage service agreements, either acquiring or delivering service. We further protect user privacy as much as is feasible–almost as much as using money protects privacy.

# Conclusions

It is widely assumed that authenticating users is a prerequisite for authorization to valuable services and that this requires each requesting user's domain to share a point of trust with each resource manager from which it requests service. Consequently people believe they need certification authorities, including commercial services such as [Verisign 1999]. The assumption is unwarranted, and certification authorities are much less important for e-business than they may seem to be.

We have solved the problem, ***"If a protected service is requested by a previously unknown consumer, how can the server decide whether or not to honor the request?"***. If the producer organization has a service agreement with an organization in which the consumer is suitably enrolled, service can safely be granted, possibly conditionally on specified further circumstances. In particular, we recognized that permission can often be derived by combining organizational enrollment of consumers with service agreements between organizations, that this information can be compactly encoded, and that certificate mechanisms based on electronic signatures can be used to combine it safely in the network. We have shown a digital network solution and variants which minimize needed human administration, which do not depend on massive infrastructure, which will be practical even for tiny servers embedded in appliances and clients in hand-held devices, and which preserve the consumer's privacy.

Formally, for a member **U** of a consuming organization **O** to get something valuable from a service **S** belonging to a producing organization **P**, it is sufficient that an agent **A(O)** of **O** and an agent **A(P)** of **P** trust each other enough to conclude a service agreement, that **A(O)** certifies that **U** belongs to **O** in some privilege category, and that **A(P)** agrees with **S** on some tickets. Each certificate and ticket can be represented by an otherwise meaningless token. Service contracts simply map certificates to tickets. Then if **U** requests from **S** and includes a certificate, **S** can ask a clearance center to determine whether or not to grant the service. No network calls to broadly accepted authorities are needed. **U** can hide its identity from **S** and **P**. Servers can deliver information and/or effect state changes.

We conjecture, but have not yet demonstrated, that our enrollment and ticket protocol also enables optimal consumer privacy.

From a historical perspective, no-one should be surprised that our problem suddenly has a practical solution. Money was invented 3000 years ago and banking 1500 years ago because, without them, commercial transactions were cumbersome when the transacting parties were far apart or not personally acquainted. When agreements can be quickly negotiated across the network, narrowly scoped tickets can often serve in lieu of money, and the network can often replace transaction services provided until now by the banking system. As digital communications become ubiquitous, rapid negotiations at a distance are becoming economical. To the extent that appliances with embedded computers become attached to the network, they will be controllable from networked personal digital assistants, e.g., computers embedded in wireless telephones, and they will need to be protected against mischievous or dangerous unauthorized manipulation. What we have described is an economical design for part of the mechansim for safe remote control.

### *Acknowledgements*

Jeff Lotspiech and Shih-Wei Luan read early versions of this report and discovered security exposures and misunderstandings of current cryptographic technology [Menezes 1997]. Günter Karjoth of the IBM Zürich Research Laboratory provided invaluable criticism of an early version of this manuscript and introduced me to the best prior work.